# Online Learning Control Strategies for Industrial Processes with Application for Loosening and Conditioning

Yue Wu [1,2]  Jianfu Cao[1]  Ye Cao[1]

(1. School of Automation, Xi'an Jiaotong University,Xi'an, 710049, China)

(2.Xinjiang Cigarette Factory, Hongyun Honghe Tobacco (Group) Co., Ltd,Urumqi, 830000, China

**Abstract:** This paper proposes a novel adaptive Koopman Model Predictive Control (MPC) framework, termed HPC-AK-MPC, designed to address the dual challenges of time-varying dynamics and safe operation in complex industrial processes. The framework integrates two core strategies: online learning and historically-informed safety constraints. To contend with process time-variance, a Recursive Extended Dynamic Mode Decomposition (rEDMDc) technique is employed to construct an adaptive Koopman model capable of updating its parameters from real-time data, endowing the controller with the ability to continuously learn and track dynamic changes. To tackle the critical issue of safe operation under model uncertainty, we introduce a novel Historical Process Constraint (HPC) mechanism. This mechanism mines successful operational experiences from a historical database and, by coupling them with the confidence level of the online model, generates a dynamic "safety corridor" for the MPC optimization problem. This approach transforms implicit expert knowledge into explicit, adaptive constraints, establishing a dynamic balance between pursuing optimal performance and ensuring robust safety. The proposed HPC-AK-MPC method is applied to a real-world tobacco loosening and conditioning process and systematically validated using an "advisor mode" simulation framework with industrial data. Experimental results demonstrate that, compared to historical operations, the proposed method significantly improves the Process Capability Index (Cpk) for key quality variables across all tested batches, proving its substantial potential in enhancing control performance while guaranteeing operational safety.

## I. Introduction

Achieving high-performance control for complex nonlinear and time-varying dynamical systems remains a pivotal challenge in modern process industries, directly impacting product quality, production efficiency, and operational safety [1], [2]. The tobacco loosening and conditioning process[3], [4], a critical stage that determines the final quality of cigarettes, epitomizes such a challenge. Its primary objective is to precisely regulate the moisture content and temperature of tobacco leaves, optimizing their physical properties for subsequent manufacturing stages[5]. However, the process is characterized by inherent complexities, including strong nonlinearities, significant time delays, multivariable coupling, and considerable variability in raw material properties [6]. These characteristics render conventional control strategies inadequate for meeting increasingly stringent quality standards. While advanced methods like fuzzy logic [7]and neural network [8]control have been explored to handle nonlinearities, they often suffer from complex rule design, poor model interpretability, or insufficient adaptability to frequent changes in operating conditions, making it difficult to achieve optimal dynamic performance while ensuring robustness[9]. Consequently, the development of a control strategy that can accurately capture nonlinear and time-varying dynamics, perform predictive optimization, and guarantee safe operation under model uncertainty remains a pressing scientific problem.

In recent years, data-driven modeling and control have offered a new paradigm for tackling these difficulties. Among these techniques, Koopman operator theory has emerged as a particularly compelling approach[10]. It provides a unique perspective by "lifting" a nonlinear dynamical system from its original state space to an infinite-dimensional function space, where the dynamics can be described by a linear Koopman operator[11]. This property offers a transformative advantage for the analysis and control of nonlinear systems. Extended Dynamic Mode Decomposition (EDMD) and its variant with control inputs (EDMDc) [12]provide a practical, purely data-driven method to compute a finite-dimensional approximation of the Koopman operator, yielding a linear predictive model that captures the essential nonlinear dynamics of the original system. Such a model is structurally transparent and can be seamlessly integrated into established linear control frameworks.

Model Predictive Control (MPC) [13], [14]is an ideal framework for leveraging such linear predictive models. By solving a Ifinite-horizon optimization problem at each sampling instant, MPC can systematically handle multivariable constraints and achieve proactive, optimal control. The combination of Koopman theory and MPC, known as Koopman-MPC[15], presents a highly promising route for controlling complex nonlinear systems. However, the direct application of a standard Koopman-MPC framework to industrial processes like tobacco loosening and conditioning faces two critical hurdles:

1.  **The Static Model-Process Mismatch:** Conventional Koopman models are identified offline in a single pass[16]. For industrial processes, dynamics can shift significantly due to factors like raw material batch changes or equipment aging [17]. A static model inevitably becomes mismatched with the real process, leading to performance degradation and potential instability. Thus, the model must possess online adaptive capabilities.

2.  **The Safety-Robustness Dilemma under Model Uncertainty:** Even with an online-updatable model, significant uncertainty exists during the initial learning phase or when the process encounters drastic, previously unseen disturbances[18]. Relying entirely on an imperfect model for optimization can lead the controller to take overly aggressive or unsafe actions, which is unacceptable in an industrial setting [19].

To address these challenges, this paper proposes a novel adaptive Koopman-MPC framework that integrates online learning with a historically-informed safety mechanism. First, to tackle the time-varying nature of the process, we employ Recursive Extended Dynamic Mode Decomposition with control (rEDMDc). By combining EDMDc with the Recursive Least Squares (RLS) algorithm[20], we construct an adaptive Koopman model whose parameters are updated in real-time as new data arrives. This endows the MPC with the ability to continuously learn and track the evolving process dynamics.

Second, to resolve the issue of safe operation under uncertainty, we introduce and integrate a novel **Historical Process Constraint (HPC)** mechanism. The rationale behind HPC is that vast amounts of historical production data contain invaluable, field-tested knowledge on how to operate safely and effectively under various conditions. The HPC mechanism is designed to transform this implicit expert knowledge into explicit, tractable constraints for the MPC optimization problem. At each control step, it identifies the most relevant successful operational instances from a historical database using a state-similarity-based K-nearest neighbors search to generate a "historical reference control." It then constructs an adaptive safety corridor around this reference, with the corridor's width dynamically linked to the confidence level of the online model. When model confidence is high, the constraints are relaxed, affording the MPC greater freedom to optimize; when confidence is low, the constraints tighten, gently guiding the control decision toward a region proven safe by historical experience. This design strikes a dynamic balance between pursuing optimal performance and ensuring robust safety, significantly enhancing the system's robustness against uncertainty.

In summary, the main contributions of this paper are threefold:

- We propose a novel adaptive Koopman-MPC framework (HPC-AK-MPC) that integrates online model identification (rEDMDc) with a safety constraint mechanism (HPC) to address the dual challenges of time-variance and model uncertainty in complex industrial processes.

- We introduce the original design of the Historical Process Constraint (HPC) mechanism, which leverages historical data mining and model confidence to provide an adaptive safety net for data-driven MPC, effectively translating implicit operational expertise into explicit control constraints.

- We systematically validate the proposed HPC-AK-MPC method on a real-world industrial application using an "advisor mode" simulation framework. The results, based on extensive industrial data, demonstrate superior performance in setpoint tracking and disturbance rejection compared to historical operations, highlighting the critical role and effectiveness of the HPC mechanism in improving control performance while guaranteeing operational safety.

## II. Preliminaries

### A. *Koopman Operator Theory*

Conventional analysis of nonlinear dynamics typically focuses on the geometric structures within the state space, such as fixed points and limit cycles. In contrast, Koopman operator theory offers an alternative perspective [19]. Instead of studying the state itself, it shifts the focus to the evolution of *observable functions* (or observables) defined on the state space. This evolution is described by a linear, albeit infinite-dimensional, operator.

Consider a discrete-time dynamical system governed by the nonlinear map $F: \mathcal{M} \to \mathcal{M}$:

$$x_{k+1} = F(x_k) \tag{1}$$

where $x_k \in \mathcal{M} \subseteq \mathbb{R}^{n_s}$ is the state of the system at time step $k$. Let $\mathcal{G}$ be a space of observable functions defined on the state space $\mathcal{M}$, where each element $g: \mathcal{M} \to \mathbb{C}$ is a mapping from the state space to the complex numbers.

The Koopman operator $\mathcal{K}: \mathcal{G} \to \mathcal{G}$ is a linear operator acting on the function space $\mathcal{G}$, defined as:

$$(\mathcal{K}g)(x) = g\big(F(x)\big) \tag{2}$$

This definition reveals the core property of the Koopman operator: it propagates the value of an observable $g$ at the current state $x_k$ to its value at the next state $x_{k+1}$. That is, if $y_k = g(x_k)$, then $y_{k+1} = g(x_{k+1}) = g(F(x_k)) = (\mathcal{K}g)(x_k)$. Crucially, although the dynamical map $F$ is nonlinear, the Koopman operator $\mathcal{K}$ is linear with respect to the function space $\mathcal{G}$:

$$\mathcal{K}(c_1 g_1 + c_2 g_2) = c_1 \mathcal{K}g_1 + c_2 \mathcal{K}g_2, \quad \forall g_1, g_2 \in \mathcal{G}, c_1, c_2 \in \mathbb{C} \tag{3}$$

This linearity implies that if we can perform a spectral decomposition of the operator $\mathcal{K}$ by finding its eigenvalues $\lambda_j$ and corresponding eigenfunctions $\phi_j(x)$ that satisfy:

$$\mathcal{K}\phi_j(x) = \lambda_j \phi_j(x) \tag{4}$$

then the time evolution of any observable $g(x) = \sum_j c_j \, \phi_j(x)$ that can be expressed as a linear combination of these eigenfunctions becomes remarkably simple:

$$g(x_k) = g\left(F^{(k)}(x_0)\right) = (\mathcal{K}^k g)(x_0) = \sum_j c_j \left(\mathcal{K}^k \phi_j\right)(x_0) = \sum_j c_j \, \lambda_j^k \phi_j(x_0) \tag{5}$$

This demonstrates that in the coordinate system formed by the Koopman eigenfunctions, the complex nonlinear dynamics are decomposed into a series of simple, linear scalar evolutions. However, the Koopman operator is typically infinite-dimensional, and its eigenfunctions are often difficult to find analytically, which limits its direct practical application.

### B. Data-Driven Approximation via EDMDc

Extended Dynamic Mode Decomposition with control (EDMDc) [12] provides a practical, data-driven method to find a finite-dimensional approximation of the Koopman operator for systems with control inputs. Given a time-series dataset of $N$ system snapshots, consisting of state-input pairs $\{x_k, u_k\}_{k=0}^N$, we first construct the corresponding data matrices. Let $\Psi_X$ be the matrix of lifted state vectors from time 0 to $N - 1$, $\Psi_{X'}$ be the matrix of their one-step-ahead counterparts from time 1 to $N$, and $U$ be the matrix of corresponding control inputs:

$$\Psi_X = [\Psi(x_0), \dots, \Psi(x_{N-1})], \quad \Psi_{X'} = [\Psi(x_1), \dots, \Psi(x_N)], \quad U = [u_0, \dots, u_{N-1}] \quad (6)$$

Assuming the truncation error $\varepsilon_k$ from the finite-dimensional approximation is statistically independent and zero-mean, the model parameter matrices $A_L$ and $B_L$ can be estimated by solving a least-squares optimization problem. Specifically, the problem aims to minimize the squared Frobenius norm of the prediction error, often with a regularization term to prevent overfitting and address data collinearity:

$$\min_{A_L, B_L} \parallel \Psi_{X'} - A_L \Psi_X - B_L U \parallel_F^2 + \lambda_{\text{reg}}(\parallel A_L \parallel_F^2 + \parallel B_L \parallel_F^2) \quad (7)$$

where $\|\cdot\|_F$ denotes the Frobenius norm, and $\lambda_{\text{reg}} > 0$ is a positive Ridge regression coefficient. To solve this, we can form an augmented data matrix $\Omega = \begin{bmatrix} \Psi_X \\ U \end{bmatrix} \in \mathbb{R}^{(n_\psi + n_u) \times N}$. The closed-form solution to the least-squares problem (7) is then given by:

$$[A_L \quad B_L] = \Psi_{X'} \Omega^\top (\Omega \Omega^\top + \lambda_{\text{reg}} I)^{-1} \quad (8)$$

Provided that the truncation error is small and the data is sufficiently exciting (i.e., the covariance matrix $\Omega \Omega^\top$ is full rank), the estimated matrices $[A_L \quad B_L]$ converge to the true projection of the Koopman operator onto the chosen subspace of observables. A key advantage of EDMDc is its computational simplicity. Once the dictionary of basis functions $\Psi$ is chosen, identifying the model parameters involves mainly matrix multiplication and inversion. However, for industrial processes like loosening and conditioning, whose dynamics change with raw material batches and environmental fluctuations, a model identified offline with fixed parameters $(A_L, B_L)$ will quickly become mismatched. This leads to performance degradation and necessitates an online parameter update mechanism.

## III. The Proposed HPC-AK-MPC Framework

### A. Online Adaptive Modeling with rEDMDc

We assume the nonlinear dynamics of the loosening and conditioning process can be described by an unknown, controlled nonlinear difference equation:

$$x_{k+1} = F(x_k, u_k, d_k) \quad (9)$$

where $x_k \in \mathbb{R}^{n_s}$ is the state vector, $u_k \in \mathbb{R}^{n_c}$ is the control input vector, and $d_k$ represents unmodeled dynamics and disturbances. The core idea is to map the original state $x_k$ into a higher-dimensional feature space $\mathcal{H}$ using a set of carefully chosen nonlinear lifting functions $\Psi: \mathbb{R}^{n_s} \to \mathbb{R}^{N_{obs}}$, where $N_{obs}$ is the dimension of the lifted space.

Ideally, if the dictionary of lifting functions $\mathcal{D}_\Psi = \{\psi_1, \ldots, \psi_{N_{obs}}\}$ spans a Koopman-invariant subspace, the system's evolution within this subspace would be linear. In practice, we select a sufficiently rich dictionary (e.g., polynomial basis functions) to approximate the dominant Koopman-invariant subspace well [20]. Based on this, we hypothesize that the system dynamics in the lifted space can be approximated by a *time-varying* linear stochastic difference equation:

$$\Psi(x_{k+1}) = A_{L,k}\Psi(x_k) + B_{L,k}u_k + w_k \tag{10}$$

where $\Psi(x_k)$ is the lifted state vector at time $k$. The matrices $A_{L,k} \in \mathbb{R}^{N_{obs} \times N_{obs}}$ and $B_{L,k} \in \mathbb{R}^{N_{obs} \times n_c}$ are the unknown, time-varying state transition and control input matrices, respectively. The term $w_k \in \mathbb{R}^{N_{obs}}$ represents the residual from the linear approximation, measurement noise, and process disturbances, which we assume to be a zero-mean random process with bounded variance.

For online parameter estimation, we consolidate the model parameters into an augmented matrix:

$$\Theta_k = [A_{L,k} | B_{L,k}] \in \mathbb{R}^{N_{obs} \times (N_{obs} + n_c)} \tag{11}$$

and define an augmented regressor vector $\Phi_k = [\Psi(x_k)^T, u_k^T]^T \in \mathbb{R}^{(N_{obs} + n_c)}$. This allows the model to be expressed compactly as:

$$\Psi(x_{k+1}) = \Theta_k \Phi_k + w_k \tag{12}$$

The task is thus to estimate the time-varying parameter matrix $\Theta_k$ online. Under standard assumptions—that the regressor sequence $\{\Phi_k\}$ is persistently exciting (PE)[21], the true parameters $\Theta_k$ vary slowly, and the process noise $w_k$ is bounded—the parameter estimation error $\widetilde{\Theta}_k = \widehat{\Theta}_k - \Theta_k$ produced by the rEDMDc algorithm is Uniformly Ultimately Bounded (UUB)[22]. This theorem guarantees that our online model will not diverge and can track the true system dynamics with a bounded error. The size of this error bound can be tuned by the forgetting factor $\lambda_f$: a smaller $\lambda_f$ provides faster tracking but is more sensitive to noise, whereas a larger $\lambda_f$ offers smoother estimates.

We employ the Recursive Least Squares (RLS) algorithm with a forgetting factor to perform the online parameter estimation. At each sampling instant $k$, upon the arrival of a new data tuple $(x_k, u_k, x_{k+1})$, the rEDMDc algorithm updates the parameter estimate $\widehat{\Theta}_{k-1}$ to $\widehat{\Theta}_k$ through the following steps:

1. **A Priori Prediction Error:**

$$e_k = \Psi(x_{k+1}) - \widehat{\Theta}_{k-1}\Phi_k \tag{13}$$

2. **Gain Vector:**

$$K_k = \frac{P_{k-1}\Phi_k}{\lambda_f + \Phi_k^T P_{k-1}\Phi_k} \tag{14}$$

3. **Parameter Update:**

$$\widehat{\Theta}_k = \widehat{\Theta}_{k-1} + e_k K_k^T \qquad (15)$$

4. **Covariance Matrix Update:**

$$P_k = \frac{1}{\lambda_f}(I - K_k \Phi_k^T)P_{k-1} \qquad (16)$$

Figure 1 illustrates the workflow of the proposed rEDMDc algorithm. The core idea is to linearize the system's dynamics by lifting the original state into a high-dimensional observable space using a set of nonlinear basis functions. The algorithm then operates in a sample-by-sample loop, employing RLS to recursively update the parameters of this high-dimensional linear model. To enhance the algorithm's adaptability to dynamic changes and ensure long-term numerical stability, two key mechanisms are incorporated: (i) an adaptive forgetting factor, which dynamically adjusts the weight of historical data based on the real-time prediction error, enabling the model to track changes quickly while effectively filtering noise during steady-state operation; and (ii) a covariance matrix reset logic, which prevents the 'covariance wind-up' problem by monitoring and bounding the trace of the covariance matrix, thereby ensuring algorithmic robustness. As a result, the algorithm outputs an updated linear model at each time step, making it an ideal choice for advanced applications like the adaptive control framework proposed herein.

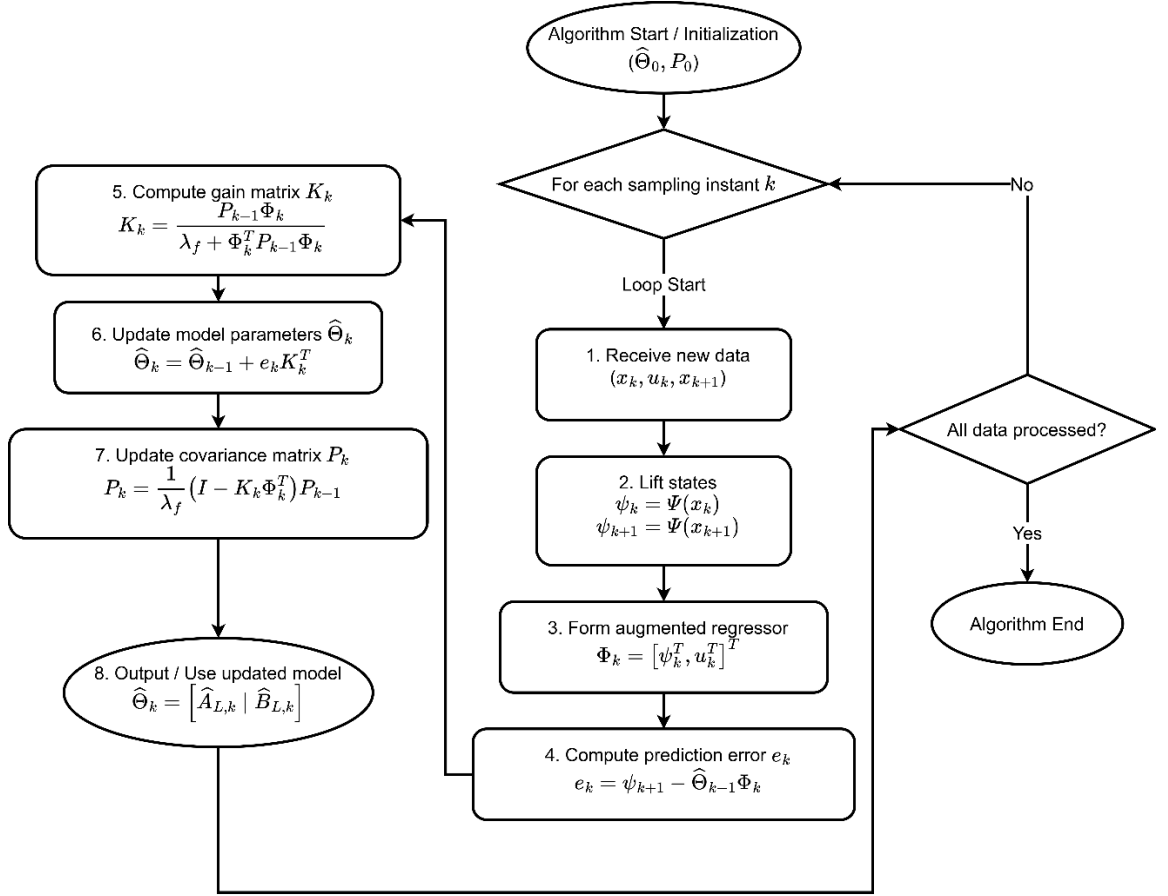

**Fig. 1. Flowchart of the rEDMDc algorithm.**

## B. Construction and Integration of Historical Process Constraints (HPC)

A key innovation of our framework is the Historical Process Constraint (HPC) mechanism, which embeds proven operational knowledge into the controller to ensure safety, especially under model uncertainty. The HPC creates a dynamic "safety corridor" for the control actions, guided by past successful operations.

Assume a historical database $\mathcal{D}$ is available, containing $N_{data}$ pairs of successfully recorded state-input data:

$$\mathcal{D} = \{(x_{\text{hist},i}, u_{\text{hist},i})\}_{i=1}^{N_{\text{data}}} \tag{17}$$

For the current state $x_k$, we first generate a *historical reference control* input, $u_{\text{ref},k}^{\text{hist}}$. This is achieved by finding the $K$ nearest neighbors to $x_k$ within the historical database $\mathcal{D}$ based on their similarity in the lifted space. The reference control is then computed as a weighted average of the control inputs corresponding to these neighbors. The procedure is detailed in Algorithm 1.

---

**Algorithm 1: Computation of Historical Reference Control**

---

Input: Current state $\Psi_{current} = \Psi(x_k)$, historical database

$\mathcal{D} = \{(\Psi_{hist,i}, u_{hist,i})\}_{i=1}^{N_{data}}$, number of neighbors K, the kernel width $\sigma_d^2$

Output: Historical reference control $u_{ref}^{hist}$
1: Lift the current state: $\Psi(x_k)$
2: For each $\Psi_{hist,i}$ in D:
3:    Calculate distance $d_i \leftarrow \parallel \Psi_{current} - \Psi_{hist,i} \parallel_2^2$
4: Find the set $I_k$ of indices of the K smallest distances
5: For each j in $I_k$:
6:    Calculate weight $\omega_j$ based on $d_j$ (e.g., $\omega_j \leftarrow \exp(-d_j/\sigma_d^2)$)
7: Compute $u_{ref}^{hist} \leftarrow u_{ref}^{hist} + \omega_j \cdot u_{hist,j}$
8: Return $u_{ref}^{hist}$

---

The boundary of the HPC should not be static but should adapt dynamically based on the confidence in the online rEDMDc model. We define a model confidence metric, $\text{Conf}_k \in [0,1]$, as follows:

$$\text{Conf}_k = \max\left(0, 1 - \frac{\text{tr}(P_k)}{\text{tr}(P_{\max})}\right) \tag{18}$$

where $\text{tr}(P_k)$ is the trace of the RLS covariance matrix $P_k$, reflecting the overall uncertainty in the parameter estimates. $P_{\max}$ is a normalization constant, typically set to the trace of the initial covariance matrix $P_0$.

The first control action computed by the MPC, $u_{k|k}$, is then constrained. For each component $j$ of the control vector, the constraint is:

$$u_{\text{ref},k}^{(j)} - \Delta u_{\text{allow},k}^{(j)} \leq u_{k|k}^{(j)} \leq u_{\text{ref},k}^{(j)} + \Delta u_{\text{allow},k}^{(j)} \tag{19}$$

where the adaptive allowable deviation $\Delta u_{\text{allow},k}^{(j)}$ is defined as:

$$\Delta u_{\text{allow},k}^{(j)} = \max\left( \left( \alpha_{\text{base}} + \beta_{\text{adapt}} \cdot \text{Conf}_k \right) \cdot \left| u_{\text{ref},k}^{(j)} \right|, \delta_{\text{abs}}^{(j)} \right) \tag{20}$$

The parameters $\alpha_{\text{base}}$, $\beta_{\text{adapt}}$, and $\delta_{\text{abs}}^{(j)}$ represent the base relative deviation, the adaptive scaling factor, and the minimum absolute deviation, respectively. Together, they define the dynamic behavior of the HPC: when model confidence is low ($\text{Conf}_k \approx 0$), the corridor is tight, guided primarily by the base deviation. As confidence grows ($\text{Conf}_k \rightarrow 1$), the corridor widens, granting the optimizer more freedom.

### *C. HPC-AK-MPC Formulation and Solution*

At each sampling time $k$, the HPC-AK-MPC controller determines the optimal control sequence $U_k = [u_{k|k}^T, \dots, u_{k+H_c-1|k}^T]^T$ over a control horizon $H_c$ by solving the following optimization problem:

$$\min_{U_k} J(U_k) = \sum_{j=1}^{H_p} \| x_{k+j|k} - x_{\text{ref},k+j} \|_Q^2 + \sum_{j=0}^{H_c-1} \left( \| u_{k+j|k} \|_R^2 + \| \Delta u_{k+j|k} \|_S^2 \right) \tag{21}$$

subject to system dynamics and constraints. Here, $H_p$ is the prediction horizon, $x_{k+j|k}$ is the predicted state at future time $k+j$, $\Delta u_{k+j|k} = u_{k+j|k} - u_{k+j-1|k}$ (with $u_{k-1|k} = u_{k-1}$), and $Q, R, S$ are positive semi-definite weighting matrices.

The state predictions are based on the most recent rEDMDc model parameters $\widehat{\Theta}_k = [\hat{A}_k | \hat{B}_k]$. The prediction for the lifted state is given by:

$$\Psi_{k+j|k} = \hat{A}_k^j \Psi(x_k) + \sum_{i=0}^{j-1} \hat{A}_k^{j-1-i} \hat{B}_k u_{k+i|k} \tag{22}$$

The original state prediction $x_{k+j|k}$ is recovered via a linear mapping $C_{\text{lift}}$ from the lifted state, i.e., $x_{k+j|k} = C_{\text{lift}} \Psi_{k+j|k}$. If the original states are included as observables in the lifting dictionary, $C_{\text{lift}}$ becomes a simple selection matrix: $C_{\text{lift}} = [I_{n_s} | 0_{n_s \times (N_{obs} - n_s)}]$.

The entire sequence of predicted states over the horizon, $X_{\text{pred},k} = [x_{k+1|k}^T, \dots, x_{k+H_p|k}^T]^T$, can be expressed as a linear function of the current lifted state $\Psi(x_k)$ and the future control sequence $U_k$:

$$X_{\text{pred},k} = \mathcal{S}_x \Psi(x_k) + \mathcal{S}_u U_k \tag{23}$$

where the prediction matrices $\mathcal{S}_x \in \mathbb{R}^{(n_s H_p) \times N_{obs}}$ and $\mathcal{S}_u \in \mathbb{R}^{(n_s H_p) \times (n_c H_c)}$ are constructed from the system matrices $\hat{A}_k$, $\hat{B}_k$, and $C_{\text{lift}}$. $\mathcal{S}_x$ represents the autonomous (or zero-input) response of the system, while $\mathcal{S}_u$ represents the forced (or zero-state) response. Their structures are derived by recursively applying (24) and are given by:

$$\mathcal{S}_x = \begin{bmatrix} C_{\text{lift}}\hat{A}_k \\ C_{\text{lift}}\hat{A}_k^2 \\ \vdots \\ C_{\text{lift}}\hat{A}_k^{H_p} \end{bmatrix}, \quad \mathcal{S}_u = \begin{bmatrix} C_{\text{lift}}\hat{B}_k & 0 & \cdots & 0 \\ C_{\text{lift}}\hat{A}_k\hat{B}_k & C_{\text{lift}}\hat{B}_k & \cdots & 0 \\ \vdots & \vdots & \ddots & \vdots \\ C_{\text{lift}}\hat{A}_k^{H_p-1}\hat{B}_k & C_{\text{lift}}\hat{A}_k^{H_p-2}\hat{B}_k & \cdots & C_{\text{lift}}\hat{A}_k^{H_p-H_c}\hat{B}_k \end{bmatrix} \tag{24}$$

### D. Conversion to a Quadratic Program (QP)

By substituting the prediction equation (23)into the cost function (21), the MPC optimization problem can be reformulated into a standard Quadratic Program (QP) of the form:

$$\min_{U_k} \quad \frac{1}{2} U_k^T H_k U_k + f_k^T U_k \tag{25}$$

where the Hessian matrix $H_k$ and the gradient vector $f_k$ are updated at each time step $k$ based on the latest model parameters and the current state:

$$H_k = 2(\mathcal{S}_u^T \bar{Q} \mathcal{S}_u + \bar{R} + \bar{S}_\Delta)$$
$$f_k^T = 2((\mathcal{S}_x \Psi(x_k) - X_{\text{ref},k})^T \bar{Q} \mathcal{S}_u - u_{k-1}^T M_{\text{prev}}^T \bar{S}_\Delta) \tag{26}$$

Here, $\bar{Q}$, $\bar{R}$, and $\bar{S}_\Delta$ are block-diagonal matrices constructed from the original weighting matrices $Q$, $R$, and $S$. As long as $R$ or $S$ is positive definite, $H_k$ is positive definite, guaranteeing a unique solution to the QP.

The constraints of the QP include physical limits (e.g., actuator saturation) and the proposed HPC. The HPC, defined in (20) applies only to the first control move $u_{k|k}$. To incorporate this into the standard QP inequality form $GU_k \leq h$, we formulate a specific constraint matrix $G_{\text{hist}}$ and vector $h_{\text{hist},k}$. The inequalities $u_{L,k} \leq u_{k|k} \leq u_{U,k}$ are stacked as $\begin{bmatrix} I \\ -I \end{bmatrix} u_{k|k} \leq \begin{bmatrix} u_{U,k} \\ -u_{L,k} \end{bmatrix}$. This is then extended to the full control sequence $U_k$ using a selection matrix:

$$G_{\text{hist}} = \begin{bmatrix} I_{n_c} & 0_{n_c \times n_c(H_c-1)} \\ -I_{n_c} & 0_{n_c \times n_c(H_c-1)} \end{bmatrix}, \quad h_{\text{hist},k} = \begin{bmatrix} u_{U,k} \\ -u_{L,k} \end{bmatrix} \tag{27}$$

The final QP to be solved at each time step $k$ combines the physical constraints ($G_{\text{phys}}U_k \leq h_{\text{phys}}$) and the HPC:

$$\min_{U_k} \quad \frac{1}{2} U_k^T H_k U_k + f_k^T U_k$$
$$\text{s.t.} \quad \begin{bmatrix} G_{\text{phys}} \\ G_{\text{hist}} \end{bmatrix} U_k \leq \begin{bmatrix} h_{\text{phys}} \\ h_{\text{hist},k} \end{bmatrix} \tag{28}$$

This is a convex QP problem that can be solved efficiently online by standard solvers.

The overall closed-loop architecture of the proposed HPC-AK-MPC controller is depicted in Figure 2. The core HPC-AK-MPC module intelligently integrates data-driven modeling with optimal control. At each step, the error between the setpoint $x_{\text{ref}}$ and the measured state $x_k$ is fed into the controller. The rEDMDc online identification block continuously uses $(x_k, u_k)$ to update the Koopman model $(\hat{A}, \hat{B})$ and its confidence metric $\text{Conf}_k$. Concurrently, the HPC module queries the historical database $\mathcal{D}$ to generate a safe reference corridor $h_{\text{hist},k}$ based on the current state and model confidence. Finally, the MPC optimizer uses the latest model and the safety constraints to compute an optimal control sequence that both drives the system towards the setpoint and respects proven operational practices. The first element of this sequence, $u_k$, is applied to the plant, completing the control loop. This architecture achieves safe and efficient control of nonlinear systems through the synergy of online model adaptation and historically-guided safety.

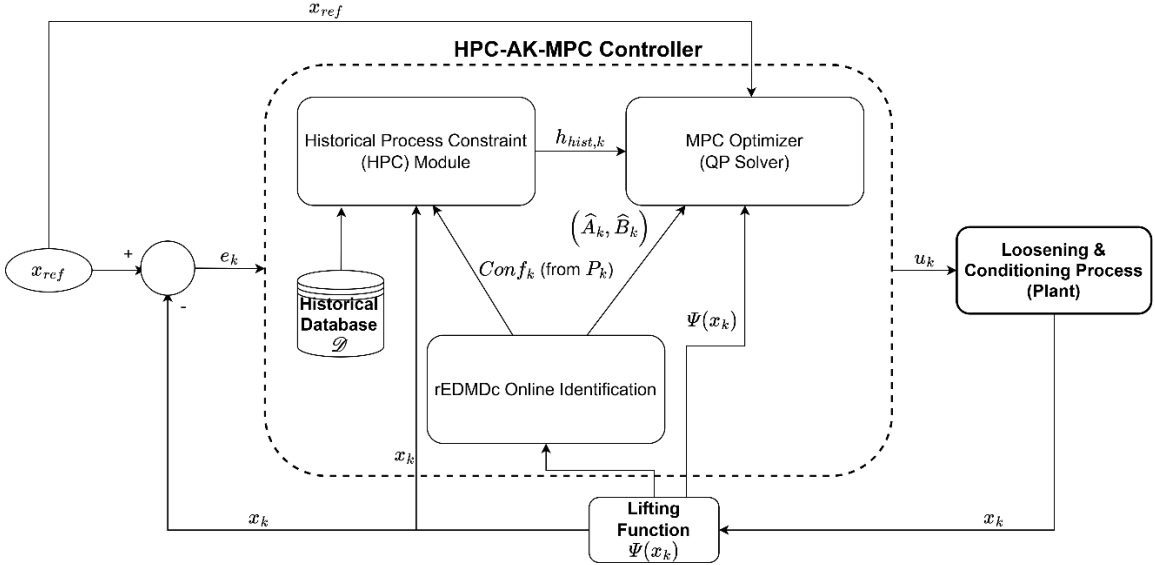

**Fig. 2. The proposed closed-loop control architecture of the HPC-AK-MPC.**

## IV. Experimental Study

### A. Industrial Process Description and Data Preprocessing

The proposed control strategy is applied to a HAUNI loosening and conditioning cylinder, a key piece of equipment in the primary processing line of a cigarette factory. The physical system consists of a large, inclined, rotating drum fitted with internal guide plates. Dried raw tobacco is fed into the higher end of the cylinder. Under the combined effects of gravity and the guide plates, the tobacco tumbles forward in a spiral motion along the inner wall. During this transport, nozzles strategically placed inside the drum spray high-temperature steam and atomized water, while a fan circulates hot air to create a specific temperature and humidity field. After approximately 5-8 minutes of intensive heat and mass exchange, the conditioned tobacco is discharged from the lower end. The loosening and conditioning process exhibits several challenging characteristics: strong nonlinearities arising from the three-phase (vapor-liquid-solid) heat and mass transfer; significant time delays due to material transport through the drum; strong coupling between variables (e.g., steam injection increases both temperature and moisture); and pronounced time-varying behavior caused by variations in raw material batches.

The primary control objective is to precisely regulate and stabilize three key quality variables (KQVs) at their process setpoints: the furnace hot air temperature, the outlet tobacco moisture content, and the outlet tobacco temperature. This constitutes a classic multi-input multi-output (MIMO) control problem. The state vector is defined as $x_k \in \mathbb{R}^3$: {Furnace Temperature (°C), Outlet Moisture (%), Outlet Temperature (°C)}, with outlet moisture being the primary control target. The input vector $u_k \in \mathbb{R}^7$ is complex, comprising five manipulated variables (MVs): {Process Throughput (kg/h), Hood Pressure (Pa), Water Flow (L/h), Steam-Water Mix Valve Opening (%), Steam Valve Opening (%)} and two key measured disturbances (DVs): {Inlet Moisture (%), Cumulative Water Added (L)}. For modeling purposes, both MVs and DVs are treated collectively as system inputs.

The dataset for this study was sourced from the Manufacturing Execution System (MES) of a cigarette factory, encompassing several months of production data for a single brand. The raw data was stored in individual Excel spreadsheets, each corresponding to a complete production batch. These files contained time-series data for all 10 variables (3 states + 7 inputs) recorded at a 1-second sampling interval.

To construct a high-quality dataset suitable for modeling and simulation, we performed the following data cleaning and preprocessing steps:

1. **Data Integration:** All batch files were read, and incomplete or excessively short batches were discarded. Valid batches were consolidated and assigned a unique numerical ID.

2. **Time Alignment:** To address minor timestamp discrepancies (on the order of seconds), all variables within each batch were time-aligned to the "Outlet Moisture" timestamp using a nearest-neighbor approach. Absolute timestamps were converted to relative time in minutes from the start of each batch.

3. **Missing Value Imputation:** Missing values were handled using linear interpolation for short gaps (up to 3 consecutive points), which is reasonable given the continuous nature of the process. Longer segments of missing data were considered unreliable and were discarded.

### B. "Advisor Mode" Simulation Framework and Experimental Setup

Evaluating a new control strategy in a real industrial setting presents significant challenges. Traditional closed-loop simulation requires a high-fidelity process model, which is difficult to obtain in a purely data-driven context where all models are derived from the same data. Furthermore, live plant experiments are often infeasible due to the non-repeatability of production conditions (e.g., raw material variations) and strict safety and quality constraints.

To overcome these obstacles, we designed an **"advisor mode" simulation framework**. The core logic of this framework is not to generate a new, independent state trajectory but to run "alongside" a real historical production batch, assessing the controller's decisions at each step. The workflow is illustrated in Figure 3 and proceeds as follows for each time step $t = 0,1, \ldots, N-1$ of a chosen historical batch:

1. **Fetch Real State:** The framework reads the actual system state, $x_{\text{actual},t}$, from the historical data log.

2. **Generate Control Advice:** The HPC-AK-MPC controller receives $x_{\text{actual},t}$. It then solves its optimization problem to generate a recommended optimal control action, $u_{\text{mpc},t}$.

3. **Predict Potential Outcome:** Using its internal, up-to-date rEDMDc model, the controller predicts what the next state, $x_{\text{pred},t+1}$, *would have been* if its recommended action $u_{\text{mpc},t}$ had been applied instead of the historical action $u_{\text{actual},t}$.

4. **Advance Time:** The simulation advances to the next time step, $t+1$, and the process repeats, using the *actual* historical state $x_{\text{actual},t+1}$ as the new measurement.

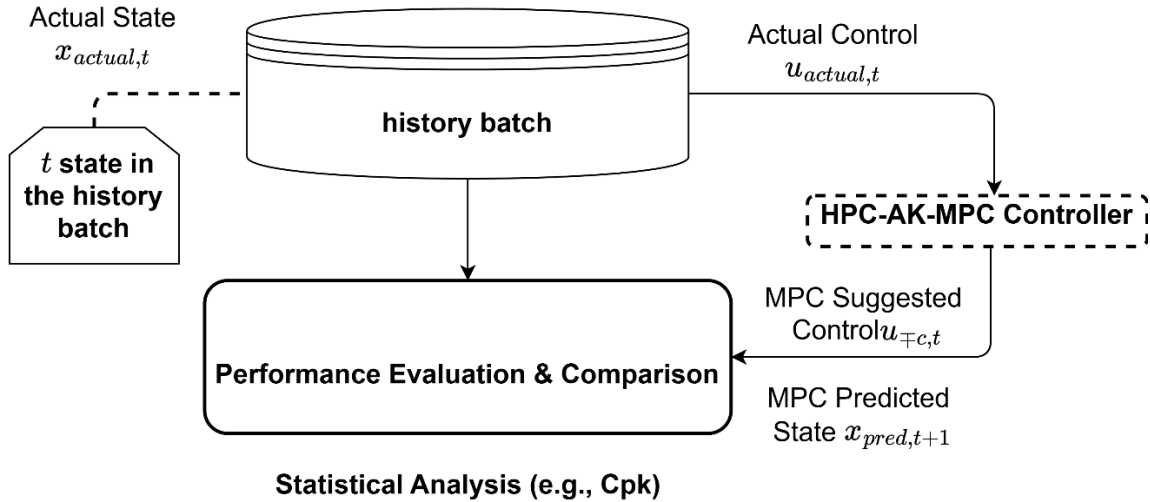

**Statistical Analysis (e.g., Cpk)**

**Fig. 3. Flowchart of the "advisor mode" simulation framework.**

The key advantage of this framework is that it allows us to evaluate the controller's decision-making quality on a completely authentic sequence of process states. It enables a direct comparison between the MPC's proposed actions and the historical plant operations, with the predicted trajectory $x_{\text{pred}}$ clearly illustrating the potential for improvement—a "what if" analysis grounded in real data.

**Experimental Parameters:**
To ensure reproducibility and fairness, the following parameters were used across all experiments:

- **rEDMDc Model:** The lifting functions consisted of a dictionary of second-order polynomials for the 3-dimensional state vector, resulting in a 10-dimensional lifted space (constant, linear, pure quadratic, and cross-product terms). The RLS algorithm used a forgetting factor of $\lambda_f = 0.995$ and a regularization coefficient of $\lambda_{\text{reg}} = 10^{-3}$. An initial parameter matrix, $\widehat{\Theta}_0$, was obtained by performing a one-time offline batch fit on a training set of 32 batches to represent the average process dynamics.

- **MPC Tuning:** The prediction and control horizons were set to $H_p = 15$ and $H_c = 7$, respectively. The state weighting matrix was $Q = \text{diag}([10,100,1])$, heavily penalizing deviations in the primary KQV, outlet moisture. The input weighting matrix was $R = 10^{-2} \cdot I_7$, and the input rate-of-change weight was $S = 0$. Physical input constraints ($u_{\min}, u_{\max}$) were set based on the 99% quantile range from the entire historical dataset, with an additional 10% margin.

- **HPC Configuration:** For simplicity in the advisor mode, the historical reference control at step $t$, $u_{\text{ref},k}^{\text{hist}}$, was set directly to the actual historical control input, $u_{\text{actual},t}$. The allowable deviation was defined as $\Delta u_{\text{allow},t} = \max(0.1 \cdot |u_{\text{actual},t}|, 0.01 \cdot \text{range}(u))$, allowing a 10% deviation from the historical value with a minimum absolute adjustment space of 1% of the variable's range.

**Performance Metric:**
Since the advisor mode does not generate a closed-loop trajectory, traditional time-integral performance metrics like IAE are not applicable. Instead, we use the **Process Capability Index (Cpk)**, a widely adopted metric in statistical process control (SPC):

$$Cpk = \min\left(\frac{\text{USL} - \mu}{3\sigma}, \frac{\mu - \text{LSL}}{3\sigma}\right) \tag{29}$$

where $\mu$ and $\sigma$ are the mean and standard deviation of the state data, and USL and LSL are the upper and lower specification limits. Cpk quantifies both the centering and the spread of the process distribution relative to its specifications, making it a gold standard for assessing process stability and quality consistency. A Cpk value greater than 1.33 is typically considered a benchmark for a capable process. We compare the Cpk calculated from the historical trajectory with that calculated from the MPC's predicted trajectory.

## C. Results and Discussion

We conducted detailed advisor mode simulations on 16 production batches not used for initial model training. The setpoints were defined based on the brand's standard as {Furnace Temp: 66.5°C, Outlet Moisture: 18.2%, Outlet Temp: 58.5°C} for testing purposes.

Figure 4 presents the simulation results for the "Furnace Temperature" on a representative test batch. The top panel shows that while the actual historical trajectory (blue line) fluctuates, the MPC's predicted trajectory (green dotted line) consistently demonstrates a strong intent to correct deviations from the setpoint (red dashed line). In the highlighted region, where the actual temperature is significantly above the target, the

MPC's prediction decisively moves downward, validating the controller's optimization logic. The bottom panel reveals the underlying control action. The MPC's proposed control for the "Water-Steam Mix Valve" (purple line) generally follows the trend of the historical operation (orange dashed line) but with more frequent and precise adjustments, leading to a more effective corrective response.

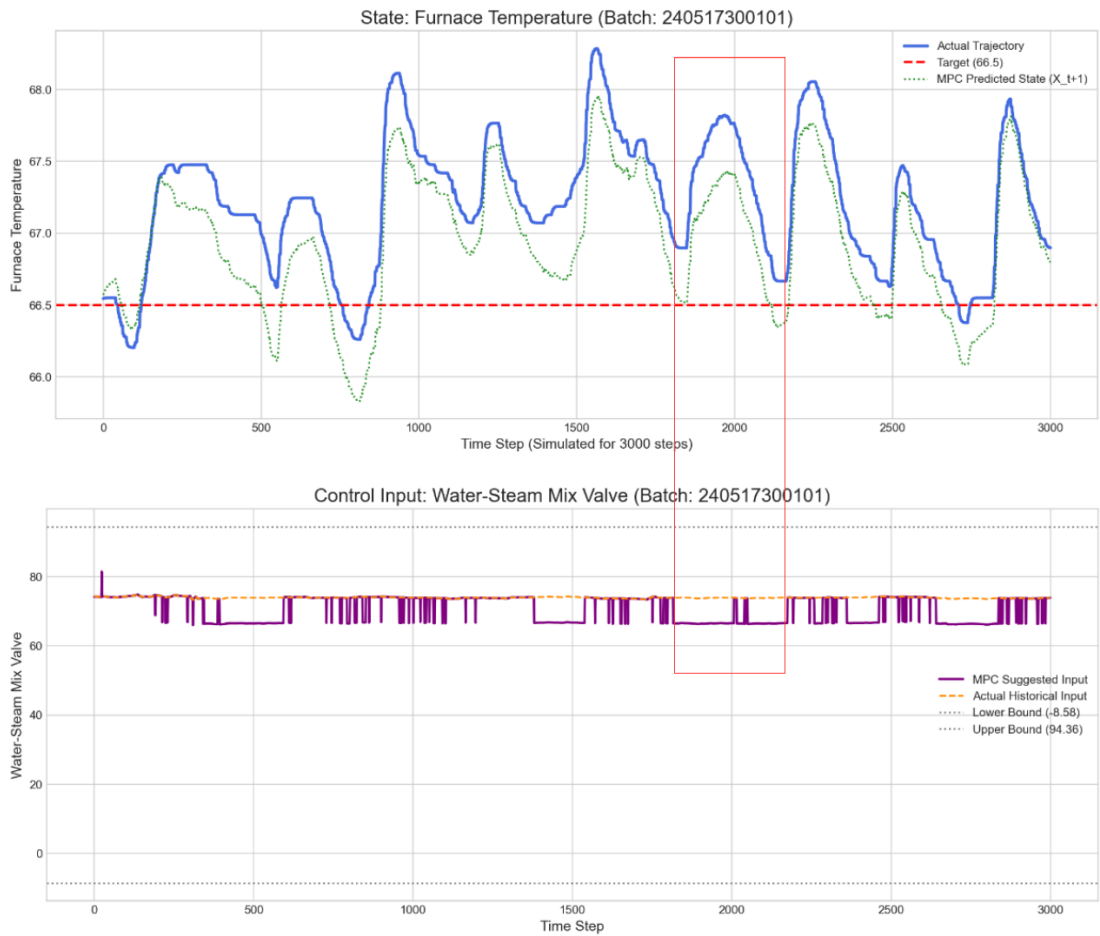

**Fig. 4. Advisor mode results for Furnace Temperature (top) and Water-Steam Mix Valve control (bottom).**

Figure 5 shows the results for the primary KQV, "Outlet Moisture." The actual trajectory (blue line) exhibits large fluctuations around the setpoint. In contrast, the MPC's predicted trajectory (green dotted line) remains much closer to the target. In the highlighted region where the actual moisture drops, the MPC's prediction clearly aims upward. This is achieved through more decisive control of the "Water Flow" (bottom panel), where the MPC (purple line) recommends clear, step-like increases to a higher flow rate to counteract the drop more aggressively than the historical operation. This again demonstrates that the MPC's optimized calculations aim for more stable and precise process control.

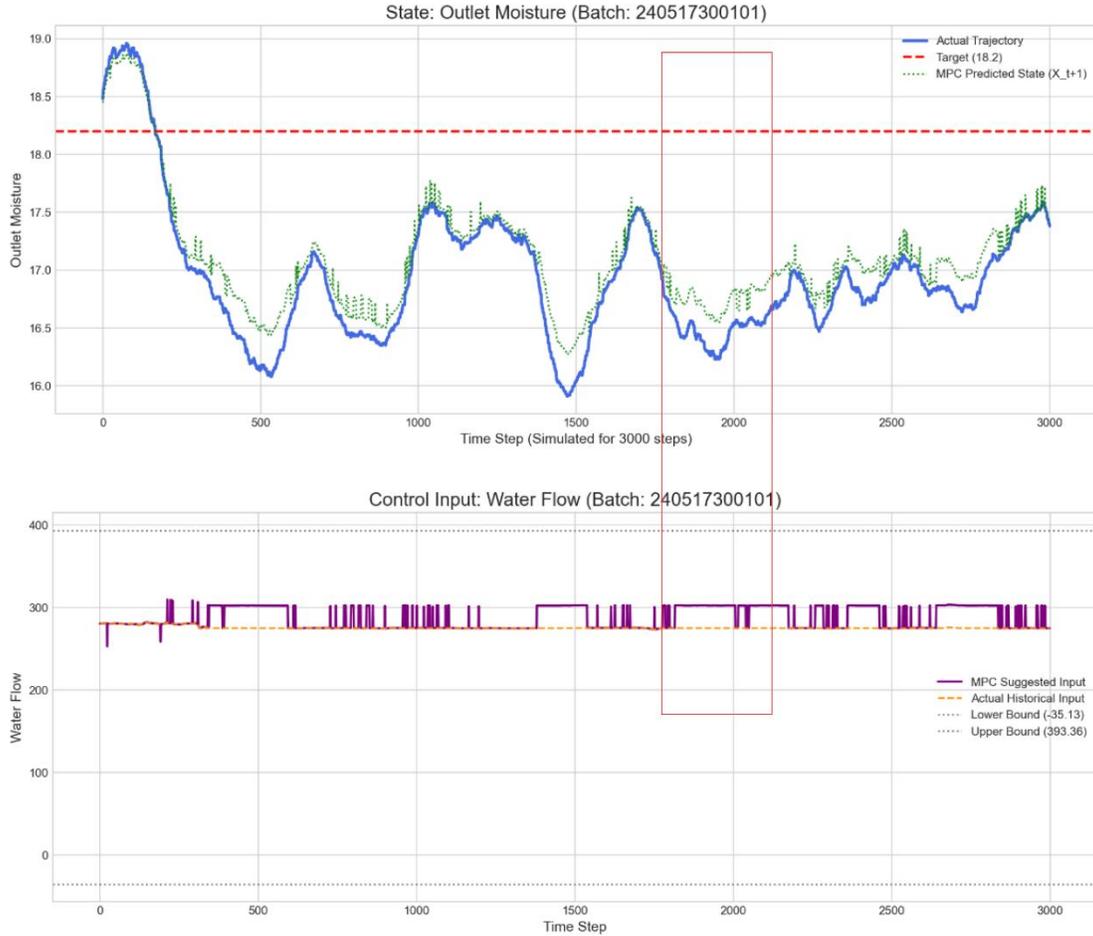

**Fig. 5. Advisor mode results for Outlet Moisture (top) and Water Flow control (bottom).**

Finally, Figure 6 provides a comprehensive comparison of the Cpk values across all 16 test batches. The box plots clearly show the potential performance improvement offered by the proposed method. For all three KQVs, the Cpk distribution of the MPC's predicted performance is significantly better than that of the historical data. For Furnace Temperature, the MPC achieves higher and more consistent capability. For Outlet Moisture, where historical control was clearly inadequate (median Cpk well below 1), the MPC shows a substantial potential for improvement. For Outlet Temperature, the MPC also offers a tighter and more centered process distribution. In summary, the MPC controller consistently demonstrates the potential to enhance or maintain the process capability of key quality variables across all tested batches.

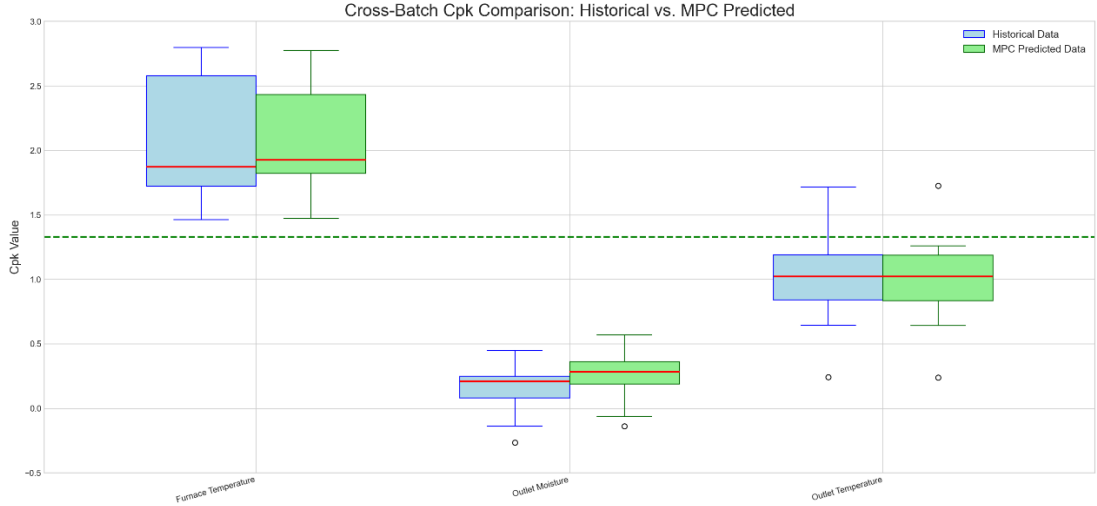

**Fig. 6. Comparison of Process Capability Index (Cpk) for historical vs. MPC-predicted performance across 16 test batches.**

## V. Conclusion

In this paper, we have successfully designed and validated a novel adaptive Koopman model predictive control framework with historical process constraints (HPC-AK-MPC) for a complex nonlinear and time-varying industrial process. The proposed method effectively addresses the critical challenges of insufficient adaptability, model-process mismatch, and the need for safe operation under uncertainty, which often hinder the application of advanced control in industrial environments. By integrating Koopman operator theory with recursive least squares, we developed an rEDMDc online identification module that enables the controller to continuously learn from real-time data and adapt its internal predictive model to track the time-varying process dynamics.

The core and most original contribution of this work is the design and integration of the Historical Process Constraint (HPC) mechanism. This mechanism translates implicit expert knowledge, embedded in historical operational data, into explicit, adaptive safety constraints for the MPC. By mining successful control examples from a historical database and dynamically linking the width of the safety corridor to the online model's confidence, the HPC creates a robust yet flexible safety net. It allows the controller to explore the optimization space freely when the model is reliable but gently guides it toward proven-safe regions when uncertainty is high. This establishes a dynamic equilibrium between the pursuit of optimal performance and the assurance of robust safety.

To validate the controller's effectiveness in an industrial context where closed-loop experiments are prohibitive, we designed and implemented an "advisor mode" simulation framework. Extensive simulations on 16 real historical production batches demonstrated the significant potential of the HPC-AK-MPC method. Compared to the actual historical operations, our controller consistently generated more precise and stable control actions, leading to predicted state trajectories that tracked setpoints more closely. This was quantitatively confirmed by a comprehensive cross-batch analysis of the Process

Capability Index (Cpk), which showed a consistent and significant improvement for all key quality variables.

In conclusion, the proposed HPC-AK-MPC framework provides a powerful and practical pathway for implementing safe, adaptive, and high-performance control on complex industrial processes, holding great promise for wide-ranging industrial applications.

## Acknowledgments

This work was supported by the Shaanxi Provincial Scientist and Engineer Research Project (Grant No. 2024QCY-KXJ-173) and the 2022 Key Science and Technology Project of Hongyun Honghe Tobacco (Group) Co., Ltd. (Grant No. HYHH2022ZN02).